\documentclass[12pt]{article}
\usepackage{epsfig}
\usepackage{amssymb}
\def\be{\begin{equation}}
\def\ee{\end{equation}}
\def\ba{\begin{array}{c}}
\def\ea{\end{array}}

\newcommand{\bea}{\begin{eqnarray}}
\newcommand{\eea}{\end{eqnarray}}

\newcommand{\kt}{\rangle}
\newcommand{\br}{\langle}
\begin{document}

\begin{center}

{\Large \bf {

%
%
%
%
%
%
%

Is $\mathcal{PT}$-symmetric quantum theory false as a fundamental
theory?

 }}

\vspace{20mm}

 {\bf Miloslav Znojil}

 \vspace{3mm}
Nuclear Physics Institute ASCR, Hlavn\'{\i} 130, 250 68 \v{R}e\v{z},
Czech Republic\footnote{The reference to the purpose of this special
issue gives me not only the opportunity for congratulations but also
for a small recollection of my first contact with Pavel Winternitz.
This happened in 1969 and was just indirect (Pavel was not in his
office in \v{R}e\v{z} at that time), mediated by Milo\v{s}
Uhl\'{\i}\v{r} (Pavel's coauthor and my teacher of quantum
mechanics). Still, this event was decisive not only for my immediate
duties (because it converted Pavel into my diploma-project leader
and - in the initial stages of development - also into my PhD thesis
supervisor) but, first of all, for all of my future, long-run
professional CV. Because the Pavel's field was mathematical physics
and because the diploma project was about relativistic quantum
S-matrix. In some sense, in spite of our subsequent long-lasting
geographical separation, I never stopped feeling inspired.}

{e-mail: znojil@ujf.cas.cz}

\vspace{3mm}



\end{center}

\vspace{5mm}

\section*{Abstract}

Yi-Chan Lee et al claim (cf. Phys. Rev. Lett. 112, 130404 (2014))
that the ``recent extension of quantum theory to non-Hermitian
Hamiltonians'' (which is widely known under the nickname of
``$\mathcal{PT}$-symmetric quantum theory'') is ``likely false as a
fundamental theory''. By their opinion their results ``essentially
kill any hope of $\mathcal{PT}$-symmetric quantum theory as a
fundamental theory of nature''. In our present text we explain that
their toy-model-based considerations are misleading and that they do
not imply any similar conclusions.

\section*{keywords}

quantum mechanics, PT symmetric representations of observables,
measurement outcomes, locality, quantum communication

\newpage

 \noindent

\section{Introduction}

At present it is still necessary to admit that even after almost
hundred years of the study of relativistic kinematics and/or of
quantum dynamics, the peaceful coexistence between our intuitive
perception of the underlying classical- and quantum-physics concepts
and principles is often fragile. This fragility dates back to the
publication of the EPR paradox \cite{EPR} and it may still be
sampled by some freshmost preprints \cite{Soucek}. In our present
paper we intend to reanalyze, critically, a re-emergence of the
conflict which we noticed in one of the very recent and very well
visible publications \cite{they}.

First of all, let us emphasize that the questions asked in
Ref.~\cite{they} are important, with possible relevance ranging from
the entirely pragmatic applications of the current quantization
principles in information theory \cite{Croke} up to pure mathematics
\cite{book}. In what follows we intend to complement the related
discussions (to be sampled, e.g., by \cite{doBro}) by a deeper
analysis and re-interpretation of some technical aspects of the toy
model as used in \cite{they}.

We may briefly summarize that our analysis will support the
affirmative answer to the question ``Could $\mathcal{PT}$-symmetric
quantum models offer a sensible description of nature?''. This
conclusion will be based, first of all, on the explicit construction
of all of the eligible physical inner products in all of the
possible related and potentially physical, ``standard'' Hilbert
space ${\cal H}^{(S)}$. In this manner, the two-parametric family of
{\em all\,} of the eligible fundamental $\mathcal{PT}$-symmetric
probabilistic interpretations of the system in question is
constructed. In full accord with the textbooks, the observables
become represented by operators which are not selfadjoint in a
``false'' Hilbert space but self-adjoint, as required, in another,
non-equivalent, ``standard'' Hilbert space. Subsequently, a few
implications of our construction will be discussed. In particular,
it will be emphasized that the conclusions of  Yi-Chan Lee et al
\cite{they} are  based on an unfortunate use of one of the simplest
but still inadequate, manifestly unphysical Hilbert spaces.

\section{Toy model}

In letter \cite{they} Yi-Chan Lee et al came with a very interesting
proposal of analysis of what happens, during the standard quantum
entangled-state-mediated information transmission between Alice and
Bob, when the Alice's local, spatially separated part $H$ of the
total Hamiltonian (say, of $H_{tot}=H\bigotimes I$ where the
identity operator $I$ represents the ``Bob's'', spatially separated
component) is chosen in the well known ${\cal PT}-$symmetric
two-level toy-model form
\begin{equation}
\label{eqn:hamiltonian}
    H=s\,
    \left ( \begin{array}{cc}
        i\sin\alpha&
1\\
        1 &-i\sin\alpha
    \end{array}
    \right ),\quad \ \ \ s,\,\alpha\in\mathbb{R}\,.
\end{equation}
The conclusions of Ref.~\cite{they} look impressive (see, i.a., the
title  ``Local {$\mathcal{PT}$} symmetry violates the no-signaling
principle''). Unfortunately, many of them (like, e.g., the very last
statement that the ``results essentially kill any hope of
$\mathcal{PT}$-symmetric quantum theory as a fundamental theory of
nature'') are based on several unfortunate misunderstandings. In
what follows we intend to separate the innovative and inspiring
aspects of the idea from some of the conclusions of Ref.~\cite{they}
which must be classified as strongly misleading and/or inadequate if
not plainly incorrect.

\subsection{$\mathcal{PT}-$symmetry.}

Our task will be simplified by the elementary nature of the
toy-model Hamiltonian $H$ of Eq.~(\ref{eqn:hamiltonian}) with
property $H{\cal PT} = {\cal PT}H$ called ${\cal PT}-$symmetry (for
the sake of clarity let us recall that one may choose here operator
$\mathcal{P}$ in the form of Pauli $\sigma_x$ matrix while
$\mathcal{T}$ may be defined simply as complex conjugation).
Secondly, our task will be also simplified by the availability of
several published reviews of the formalism (let us call it ${\cal
PT}-$symmetric Quantum Mechanics, PTQM)  and, in particular, of its
recent history of development (let us recall, say, its two most
exhaustive descriptions as provided by Refs.~\cite{Carl,ali}).

Incidentally, it is extremely unfortunate that the latter two PTQM
summaries remained, obviously, unknown to or, at least, uncited by,
the authors of letter \cite{they} (for the sake of brevity let us
call this letter ``paper I'' in what follows). Otherwise, the
authors of paper I would be able to replace their first and already
manifestly incorrect description of the birth of the formalism (in
fact, the first sentence of their abstract which states that in
1998, ``Bender \emph{et al.}~\cite{Bender1998} have developed
$\mathcal{PT}$-symmetric quantum theory as an extension of quantum
theory to non-Hermitian Hamiltonians'') by some more appropriate
outline of the history. Reminding the readers, e.g., that for the
majority of active researchers in the field (who are meeting, every
year, during a dedicated international conference \cite{phhqp}) the
presently accepted form of the $\mathcal{PT}$-symmetric quantum
theory has only been finalized, roughly speaking, after the
publication of the ``last erratum'' \cite{erratum} in 2004.

Naturally, even the year 2004 was not the end of the history since
during 2007, for example, the description of the so called
$\mathcal{PT}$-symmetric brachistochrone~\cite{Bender2007} moved a
bit out of the field and had to be followed by the thorough
(basically, open-system-related) re-clarification of the concept
(cf.~\cite{alibrach} and also, a year
later,~\cite{Gunther2008,Gunther2008a}). During the same years also
the methods of extension of the intrinsically non-local PTQM
formalism to the area of scattering experiments were developed
\cite{Cannata,Jones,scatt}.

\subsection{Eligible physical inner products.}

Unfortunately, the authors of paper I have missed the latter
messages. Having restricted their attention solely to the
brachistochronic quantum-evolution context of the initial
publication~\cite{Bender2007} they remained unwarned that in this
context the role of the generator $H$ may be twofold. It is being
used {\em either} in the {\em unitary} quantum-evolution context of
Refs.~\cite{Carl,ali} (cf. also the highly relevant, cca fifteen
years older review paper~\cite{Geyer}) {\em or} in application to
the open quantum systems.

In the latter case one is allowed to speak just about a {\em
non-unitary}, truly brachistochronic quantum evolution {\em within a
subspace} of a ``full'' Hilbert space of
states~\cite{Gunther2008,Gunther2008a}. Naturally, the quantum world
of the above-mentioned Alice cannot belong to such a category. In
other words, ``her'' Hamiltonian (\ref{eqn:hamiltonian}) {\em must
necessarily} be made self-adjoint. According to the standard theory
(briefly reviewed also in our compact review \cite{SIGMA}), this
should be made via a replacement of the ``friendly but false''
Hilbert space ${\cal H}^{(F)}$ (chosen, in paper I, as ${\cal
H}^{(F)}\ \equiv \ \mathbb{C}^2$ for model (\ref{eqn:hamiltonian}))
by another, ``standard, sophisticated'' Hilbert space ${\cal
H}^{(S)}$ which only differs from ${\cal H}^{(F)}$ in its use of a
{\em different inner product} between its complex two-dimensional
column-vector elements $|a\kt = (a_1,a_2)^T$ and $|b\kt =
(b_1,b_2)^T$.

The usual and ``friendly'', $F-$superscripted inner product
 \be
 \br a|b\kt = \br a|b\kt^{(F)} = \sum_{i=1,2} a_i^*\,b_i
 \ee
defines the Hilbert-space structure in the false and manifestly
unphysical, ill-chosen and purely auxiliary friendlier space ${\cal
H}^{(F)}$. Thus, what is now required by the PTQM postulates is an
introduction of a {\em different}, $S-$superscripted product
 \be
 \br a|b\kt^{(S)} = \sum_{i,j=1,2} a_i^*\,\Theta_{ij}\,b_j
 \label{newin}
 \ee
containing an {\em ad hoc} (i.e., positive and Hermitian
\cite{Geyer}) ``Hilbert-space-metric'' matrix $\Theta=\Theta^{(S)}$.
Precisely this enables us to reinterpret our given Hamiltonian $H$
with real spectrum as living in a manifestly physical, new Hilbert
space ${\cal H}^{(S)}$. Naturally, one requires that such a
Hamiltonian generates a unitary evolution in the correct, physical
Hilbert space ${\cal H}^{(S)}$ or, in mathematical language, that it
becomes self-adjoint with respect to the upgraded inner product
(\ref{newin}).

\section{Physics}

\subsection{Admissible probabilistic interpretations of the model.}

For our two-dimensional matrix model (\ref{eqn:hamiltonian}) the
latter condition proves equivalent to the set
 \be
 H^\dagger\,\Theta=\Theta\,H
 \ee
of four linear algebraic equations with general solution
 \be
 \Theta=
 \label{eqn:metric}
    a^2\,
    \left ( \begin{array}{cc}
    1&   u- i\sin\alpha
 \\
         u+ i\sin\alpha &1
    \end{array}
    \right ),\quad \ \ \ a,\,u,\,\alpha\in\mathbb{R}\,, \ \ \ |u|<|\cos \alpha|.
\end{equation}
Any choice of admissible parameter $u$ is easily shown to keep this
metric (as well as its inverse) positive. Thus, the reason why the
parameter $\alpha$ was called ``the non-Hermiticity of $H$'' in
Ref.~\cite{Gunther2008} is purely conventional, based on a tacit
assumption that one speaks, say, about an open quantum system. On
the contrary, once we restrict our attention to the world of Alice
(who {\em must} live in the {\em physical} Hilbert space ${\cal
H}^{(S)}$), we {\em must} speak about the {\em unitarily} evolving
quantum states and about the relevant generator
(\ref{eqn:hamiltonian}) which is, by construction, {\em Hermitian}
inside any pre-selected physical Hilbert space, given by our choice
of the free parameter $u$.

For this reason the calculation of what, according to paper I, ``Bob
will measure using conventional quantum mechanics'' must be again
performed in the physical Hilbert space. In particular, the trace
formulae as used in paper I are incorrect and must be complemented
by the pre-multiplication of the bra vectors by the ``shared''
metric from the right, $\br \psi_f| \ \to \
\br\psi_f|\tilde{\Theta}$ (in the most elementary scenario one could
simply recall Eq.~(\ref{eqn:metric}) and choose $\tilde{\Theta}=
\Theta \bigotimes I$).

\subsection{Observables.}

Many of the related comments in paper I (like, e.g., the statement
that ``These states [given in the unnumbered equation after Eq.
Nr.~(2)] are not orthogonal to each other in conventional quantum
theory'') must be also modified accordingly. The point is that the
non-orthogonality of the eigenstates of $H$ in the manifestly
unphysical Hilbert space ${\cal H}^{(F)}$ is entirely irrelevant. In
contrast, what remains decisive and relevant is that, in the words
of paper I, ``when $\alpha=\pm\pi/2$, they become the same state,
and this is the $\mathcal{PT}$ symmetry-breaking point''. Indeed,
one easily checks that in such an ``out-of-theory'' limit (towards
the so called Kato's exceptional point \cite{Kato}) the metric (and,
hence, the physical Hilbert space) ceases to exist.

One has to admit that the currently accepted PHQP terminology is a
bit unfriendly towards newcomers. Strictly, one would have to speak
about the Hermiticity of any two-by-two matrix observable $\Lambda$,
i.e., equivalently, about the validity of the necessary Hermiticity
condition in physical space,
 \be
 \Lambda^\dagger\,\Theta=\Theta\,\Lambda\,.
 \label{druha}
 \ee
Naturally, this condition can only be tested {\em after} we choose a
definite form of the metric (\ref{eqn:metric}), i.e., in our toy
model, after we choose the inessential scale factor $a^2>0$ and the
essential metric-determining parameter $u$ in
Eq.~(\ref{eqn:metric}).

It is worth adding that in order to minimize possible confusion the
authors of the oldest review paper \cite{Geyer} recommended that,
firstly, whenever one decides to work with a nontrivial (sometimes
also called ``non-Dirac'') metric $\Theta\neq I$, the natural
Hermiticity condition in the ``hidden'' physical space should be
better called ``quasi-Hermiticity''. Secondly, they also recommended
that having a Hamiltonian, there may still be reasons for our
picking up a suitable candidate $\Lambda$ for another observable in
advance. Then, equation (\ref{druha}) would acquire a new role of an
additional phenomenological constraint imposed upon the metric.

Incidentally, in the PTQM context the latter idea found its
extremely successful implementation in which one requires that the
second observable $\Lambda$ represents a charge of the quantum
system in question. It is rather amusing to verify that such a
specific requirement (called, sometimes, ${\cal PCT}$ symmetry
\cite{Carl}) would remove all of the ambiguities from the metric of
Eq.~(\ref{eqn:metric}) simply by fixing the value of $u=0$ as well
as of $a^2=1/\cos \alpha$.

\section{Conclusions}

We are now prepared to return to the two key PTQM assumptions as
formulated in paper I. Their main weakness is that they use the
concept of the physical Hilbert space (i.e., in essence, of the
unitarity of evolution) in a very vague manner. One should keep in
mind that even in the phenomenologically extremely poor
two-dimensional toy models the predictions and physical content of
the theory are very well understood as given not only by the
generator of evolution $H$ but also by the second observable
$\Lambda$ (say, charge - for both, naturally, we require that the
spectrum is real). Thus, what can be measured in the model is the
energy and, say, charge. In other words, the theory does not leave
any space for any kind of coexistence between different
``conventional'' metrics and/or between different normalization
conventions (i.e., typically, for the simultaneous use of different
parameters $u$  in Eq.~(\ref{eqn:metric})). At the same time, in the
light of paper \cite{scatt} on the PTQM-compatible unitarity of the
scattering, the PTQM theory still leaves space for a consistent
implementation of the important phenomenological concepts like
locality, etc.

The concluding remarks of paper I about a conjectured  ``trichotomy
of possible situations'' must be thoroughly reconsidered. Keeping in
mind the necessary separation of alternative PTQM-related problems
and eliminating, first of all, any mixing between the two well
defined categories, viz., of the quantum models characterized by the
unitary and/or non-unitary evolution. Definitely, the theories of
the PTQM type did not exhaust their potentialities yet. It is truly
impossible to agree with the final statement of paper I that its
``results essentially kill any hope of $\mathcal{PT}$-symmetric
quantum theory as a fundamental theory of nature''.


\subsection*{Acknowledgements}

The work on the project was supported by the Institutional Research
Plan RVO61389005 and by GA\v{C}R Grant Nr. 16-22945S.

\end{document}